\documentclass[5p,twocolumn]{elsarticle}

\usepackage{lineno,hyperref}
\modulolinenumbers[5]

\journal{Chaos, Solitons and Fractals}

\usepackage{graphicx,epstopdf}
\usepackage{epsfig}
\usepackage{subfigure}
\usepackage{amsmath,amssymb,amstext}
\usepackage[export]{adjustbox}
\usepackage{float}
\usepackage{comment}
\usepackage{graphicx}
\usepackage{xcolor}
\usepackage{hyperref}

\usepackage[numbers]{natbib}
\usepackage{comment}
\usepackage{mathtools}
\usepackage[export]{adjustbox}
\usepackage{overpic}
\usepackage{mathscinet}
\usepackage{bm}
\usepackage{amsthm}
\usepackage{cleveref} 
\usepackage{nameref} 

\usepackage{float}
\usepackage{dcolumn}
\usepackage{bm}
\usepackage{bigints}
\usepackage{xcolor}

\usepackage{physics}
\usepackage{mleftright}
\usepackage{nicefrac}
\usepackage{enumitem}

\newcommand{\Eq}[1]{Eq.~\eqref{#1}}
\newcommand{\Eqb}[2]{Eqs.~(\ref{#1},~\ref{#2})}
\newcommand{\refpan}[2]{\hyperref[#1]{\ref*{#1}.#2}}

\newcommand{\bx}{\ensuremath{\mathbf{x}}}
\newcommand{\by}{\ensuremath{\mathbf{y}}}

\newcommand{\ullK}{\underline{\underline{\mathbf{K}}}}

\newcommand{\avg}[1]{\left\langle{#1}\right\rangle}

\newcommand{\x}{\mathbf{x}}
\newcommand{\y}{\mathbf{y}}

\begin{document}

\title{Anticipated synchronization in systems with distributed delay}

\author{David Ortiz del Campo}
\ead{davidortiz@ifisc.uib-csic.es}
\author{Tobias Galla}
\ead{tobias.galla@ifisc.uib-csic.es}
\author{Ra\'ul Toral}
\ead{raul@ifisc.uib-csic.es}
\affiliation{Instituto de Fisica Interdisciplinar y Sistemas Complejos IFISC (CSIC-UIB), Campus UIB, 07122 Palma de Mallorca, Spain.}

\date{\today}

\begin{abstract}

Anticipated synchronisation occurs when a driven dynamical system synchronises with the future state of the driver system to which it is unidirectionally coupled. Previous theoretical and experimental studies have focused on setups with a single delay time in the coupling term, for which exact anticipation can arise as a solution. Here we extend this framework to configurations with distributed delay times. Our main result is that, for a given delay distribution, approximate anticipated synchronisation can emerge over a range of coupling strengths. We analyse this phenomenon analytically for systems of linear oscillators, where we identify simple cases exhibiting exact synchronisation--up to a constant amplitude factor. Numerical simulations of nonlinear chaotic systems reveal stable forms of approximate anticipated synchronisation.
\end{abstract}
\maketitle

\section{Introduction}

Anticipated synchronisation (AS) is a counter-intuitive form of synchronisation in which a dynamical system aligns with the future state of another system to which it is coupled. Introduced by Voss~\cite{Voss_2000,Voss_2001b}, different forms of AS have been demonstrated in a variety of theoretical models and experimental settings, including linear and nonlinear systems~\cite{Calvo_2004,Masoller_2001,Hernandez_Garcia_2002, Voss_2016}, chaotic dynamics~\cite{Voss_2001a,Shahverdiev_2002,Pyragiene_2015}, excitable media~\cite{Ciszak_2003}, electronic circuits~\cite{voss2002real}, electrochemical systems ~\cite{PhysRevE.111.064206}, optical systems~\cite{Masoller_2001_lasers, liu2002experimental,Sivaprakasam_2001,Rees_2003,Tang_2003,Wang_2005}, neuronal circuits ~\cite{Ciszak_2003,Toral_2003,Ciszak_2004, Matias_2011, Matias_2013, Sausedo_Solorio_2014, Montani_2015, Mirasso_2017, Dalla_Porta_2019}, and manual tracking~\cite{ Stepp_2017}. AS has been proposed as a mechanism to control dynamical systems~\cite{Ciszak_2009,Mayol_2012,Zamora-Munt_2014}, and for estimating parameters of chaotic systems~\cite{Wei_2010}. More recently, AS has also been shown to occur in spatially extended and spatiotemporal chaotic systems~\cite{Ciszak_2015}.

\begin{figure}[t]
	\centering
	\includegraphics[width=.9\linewidth]{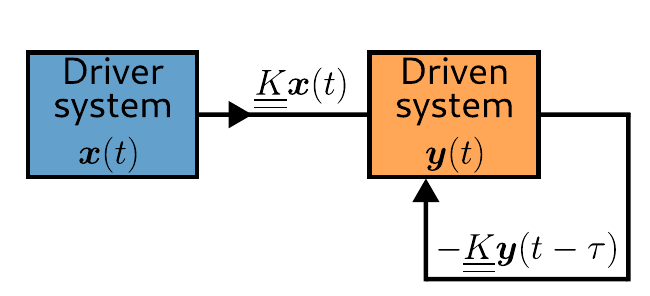}
	\caption{Illustration of the driver $\bx(t)$ and the driven system $\by(t)$. The driver operates autonomously. The driven system is influenced unidirectionally by the driver through a term $K\mathbf{x}$. There is also a delay term $K\mathbf{y}(t-\tau)$ feeding into the dynamics of $\by$ at time $t$. As explained in the text, we refer to $\ullK \left[ \mathbf{x}\mleft( t \mright) - \mathbf{y}\mleft( t - \tau \mright) \right]$ as the `delayed coupling term'.}
	\label{fig:driver_driven}
\end{figure}

In the present work we focus on a setup in which AS is achieved by coupling two identical systems in a \emph{driver--driven} configuration~\cite{Voss_2000, Voss_2001b}. The \emph{driver} system, $\mathbf{x}(t)$, evolves autonomously, while the \emph{driven} system, $\mathbf{y}(t)$, receives a unidirectional input from the driver and includes a delayed self-feedback term, as illustrated in Fig.~\ref{fig:driver_driven}. 

For suitable values of the coupling strength and delay time, the dynamics converges to the anticipated manifold
\begin{equation}
\mathbf{y}(t) = \mathbf{x}(t+\tau),
\label{eq:AS_manifold}
\end{equation}
so that the driven system anticipates the evolution of the driver by a time $\tau$. Remarkably, this anticipation occurs without altering the dynamics of the driver system, and despite the fact that both systems evolve according to strictly causal equations.

Nearly all previous studies of anticipated synchronization assume that the coupling between the driver and the driven systems involves a single fixed delay. While this assumption is convenient, it is often unrealistic: delays may fluctuate, be not exactly known, or arise from multiple interaction channels with different transmission times. This naturally leads to the concept of \emph{distributed delay}, in which the delayed feedback acting on the driven system is described by a distribution of delays rather than a single delay time. While most setups assume a constant delay, some investigations have introduced time-dependent delays. In particular periodically varying delays were implemented in \cite{Ambika_2009}.

Our main objective in this paper is to investigate if and when anticipated synchronization can occur in systems with distributed delay coupling, and to characterize how the presence of a delay distribution affects anticipation. In particular, we ask if exact or approximate anticipation can be observed, what the effective anticipation time is, and how distributed delay modifies the stability and qualitative features of the synchronized state. 

We address these questions using a combination of analytical and numerical methods. We gain insight into the role of distributed delay by studying linear systems, specifically coupled damped harmonic oscillators. A complete analytical solution can then be found, and it can be used to identify conditions under which anticipation emerges. For nonlinear systems, including chaotic dynamics, an analytical solution does not appear to be possible. While we do not find exact anticipated synchronisation in numerical simulations, approximate anticipated synchronization can still arise. In those cases we find that the anticipation time is determined by the delay distribution and remains close to the average delay in the coupling between the driver and driven systems. Our results show that anticipated synchronization is a robust phenomenon that extends beyond the idealized case of single-delay coupling.

The remainder of the paper is organised as follows. In Sec.~\ref{sec:setup} we describe the general setup of the driver and driven systems. In order to develop further insight we then focus on linear systems in Sec.~\ref{sec:linear}, building for example on \cite{Calvo_2004}. In particular we provide an analytical characterisation of AS in systems of harmonic oscillators subject to coupling with distributed delay. In Sec.~\ref{sec:chaotic} we then show numerically that approximate AS can occur in chaotic systems with distributed delay in the coupling terms. Finally, we present our conclusions in Sec.~\ref{sec:conclusions}.

\section{General setup: Driver-driven systems with delay}\label{sec:setup}
\subsection{Coupling with a fixed delay}

We consider two identical dynamical systems coupled in a unidirectional driver--driven configuration. The state of each system is described by an $n$-dimensional vector of real-valued variables. The driver system $\mathbf{x}(t)=(x_1(t),\cdots,x_n(t))$ evolves autonomously according to
\begin{equation}
\dot{\mathbf{x}}(t) = \mathbf{f}[\mathbf{x}(t)],
\label{eq:driver}
\end{equation}
where $\mathbf{f}$ defines the intrinsic dynamics. 

The driven system $\mathbf{y}(t)=(y_1(t),\ldots,y_n(t))$ obeys the same intrinsic dynamics, but in addition receives an instantaneous positive input from the driver together with a delayed negative self-coupling term~\cite{Voss_2000,Voss_2001a,Voss_2001b},
\begin{align}
\dot{\mathbf{y}}(t) &= \mathbf{f}[\mathbf{y}(t)] + 
\ullK \left[\mathbf{x}(t) - \mathbf{y}(t-\tau) \right].
\label{eq:fixed_delay_system}
\end{align}
where $\ullK$ is an $n \times n$ coupling matrix and $\tau$ is a fixed delay time. 

The system of Eqs.~(\ref{eq:driver} \ref{eq:fixed_delay_system}) admits the exact anticipated synchronization solution $\mathbf{y}(t) = \mathbf{x}(t+\tau)$ [\Eq{eq:AS_manifold}] for which the coupling term vanishes identically and the driven system anticipates the trajectory of the driver by a time $\tau$. The stability of such solution for suitable choices of $\ullK$ and $\tau$ has been studied extensively in a variety of systems~\cite{Voss_2000,Voss_2001a,Voss_2001b,Calvo_2004,Ciszak_2005}.

\subsection{Coupling with distributed delay\label{sec:CoupDistrDelay}}

We now generalize the coupling scheme to the case of distributed delay where the self-feedback of the driven system is governed by a normalized delay distribution $g(\tau)$, such that the full dynamics reads
\begin{align}
\dot{\mathbf{x}}(t) &= \mathbf{f}[\mathbf{x}(t)], \nonumber\\
\dot{\mathbf{y}}(t) &= \mathbf{f}[\mathbf{y}(t)] +
\ullK \left[
\mathbf{x}(t) -
\int_0^\infty d\tau\, g(\tau)\,\mathbf{y}(t-\tau)
\right].
\label{eq:distributed_delay_system}
\end{align}
Distributed delay naturally arises in situations where the delay is uncertain, fluctuates in time, or results from the coexistence of multiple interaction pathways with different transmission times. The fixed-delay case with delay time $\tau_s$ is recovered by choosing $g(\tau)=\delta(\tau-\tau_s)$.

In general, systems with distributed delay do not admit exact anticipated synchronization solutions of the form of \Eq{eq:AS_manifold}. However, as we will demonstrate, the driven signal is often found to be similar to that of the driver up to a time shift,
\begin{equation}\label{eq:app_ant}
\mathbf{y}(t) \approx \mathbf{x}(t+\tau^\ast).
\end{equation}
Here $\tau^\ast$ is an effective anticipation time. In such cases we speak of \emph{approximate anticipated synchronization}. To quantify this similarity we use cross-correlation measures between corresponding components of $\mathbf{x}(t)$ and $\mathbf{y}(t)$, 
\begin{equation}
 \rho_{x_i, y_j}(\Delta t) = \frac{\avg{\left[ x_i(t) - \avg{x_i}_T \right] \cdot \left[ y_j(t+\Delta t) - \avg{y_j}_T \right]}_{T-\Delta t}}{\sigma_{x_i} \sigma_{y_j}},
\end{equation}
where we define
\begin{eqnarray}
 \avg{f}_T &\equiv& \frac{1}{T}\int_{0}^{T}\dd{t}f(t), \\
 \sigma_f^2 &\equiv& \avg{\left(f - \avg{f}_T\right)^2}_T,
\end{eqnarray}
for an arbitrary function $f(t)$.

These quantities typically converge for sufficiently large values of $T$, provided the associated averages are well defined (e.g., in the absence of divergences in the relevant functions).

\section{Linear systems: Damped harmonic oscillator}\label{sec:linear}

\subsection{Setup and AS solution}\label{sec:single_oscillator}
We consider a driver--driven system of identical damped harmonic oscillators subject to distributed delay coupling. The driver system $\x=(x_1,x_2)$ is governed by
\begin{align}
\dot{x}_1(t) &= x_2(t), \nonumber\\
\dot{x}_2(t) &= -\Omega^2 x_1(t) - 2 b x_2(t),
\label{eq:driver_osc}
\end{align}
while the driven system $\y=(y_1,y_2)$ evolves according to
\begin{align}
		\dot{y}_1(t) =& y_2(t) + K_1 \left[ x_1(t) - \int_0^\infty\dd{\tau}g(\tau)y_1(t-\tau) \right],\nonumber \\
 \dot{y}_2(t) =& - \Omega^{2} y_1(t) - 2by_2(t),\nonumber \\
 &+ K_2 \left[ x_2(t) - \int_0^\infty\dd{\tau}g(\tau)y_2(t-\tau) \right].\label{eq:driven_osc}
\end{align}
Here $\Omega>0$ denotes the natural frequency, $b\ge0$ the damping coefficient, and $K_1$ and $K_2$ are the coupling strengths for the first and second components respectively.

In the underdamped regime $b<\Omega$, the driver exhibits exponentially decaying oscillations 
\begin{equation}\label{eq:SolutionX1Master}
 x_1(t) = Ae^{-bt}\cos\mleft( \Omega_0 t + \phi \mright),
\end{equation}
with frequency $\Omega_0=\sqrt{\Omega^2-b^2}$, and integration constants $A$ and $\phi$. We seek solutions for the driven system in \Eq{eq:driven_osc} for which the delayed coupling terms vanish identically, i.e. where for all times $t$
\begin{equation}
x_i(t)=\int_0^\infty d\tau\, g(\tau)\, y_i(t-\tau),
\qquad i=1,2.
\label{eq:vanishing_coupling}
\end{equation}
The solution can be found in the Fourier space. From \Eqb{eq:driver_osc}{eq:SolutionX1Master} we find
\begin{eqnarray}\label{eq:SolutionX1MasterFourier}
 \tilde{x}_1(\omega) &=& \frac{A}{2} \left[ \delta \mleft( ib + \Omega_0 - \omega \mright) + \delta \mleft( ib - \Omega_0 - \omega \mright) \right],\\
\tilde{x}_2(\omega) &=&i\omega \tilde{x}_1(\omega),
\end{eqnarray}
where we use the notation $\tilde{f}(\omega)$ for the Fourier transform of a function $f(t)$. Using the convolution theorem in \Eq{eq:vanishing_coupling} we find $\tilde{y}_i(\omega)=\tilde{x}_i(\omega)/\tilde{g}(\omega)$. Inverting the Fourier transform we then obtain,
\begin{equation}
y_i(t)=\beta\, x_i(t+\tau^\ast),
\qquad i=1,2,
\label{eq:AS_solution}
\end{equation}
where the effective anticipation time $\tau^\ast$ and the amplitude factor $\beta$ are given by
\begin{align}
\beta &= e^{b \tau^\ast}\left|\int_0^\infty d\tau\, g(\tau)\, e^{b\tau} e^{i\Omega_0 \tau}\right|^{-1}, \label{eq:beta}\\
\tau^\ast &= \Omega_0^{-1}
\arctan\!\left(
\frac{\int_0^\infty d\tau\, g(\tau)\, e^{b\tau}\sin(\Omega_0\tau)}
{\int_0^\infty d\tau\, g(\tau)\, e^{b\tau}\cos(\Omega_0\tau)}
\right).
\label{eq:tstar}
\end{align}
If this solution is realised the driven oscillator therefore anticipates the trajectory of the driver by a time $\tau^\ast$. In addition, the amplitude of the driven system is multiplied by a factor $\beta$ relative to that of the driver. This implies that amplification of the driven response relative to the driver can occur. For a single fixed delay, $g(\tau)=\delta(\tau-\tau_s)$, one recovers exact anticipated synchronization with $\tau^\ast=\tau_s$ and $\beta=1$. 

\subsection{Stability analysis and types of dynamic behaviour}
To assess whether the anticipated solution is dynamically stable, we define the coupling mismatches \makebox{$\mathbf{\Delta}(t)=(\Delta_1(t),\Delta_2(t))$} as
\begin{equation}
\mathbf{\Delta}(t)=\x(t)-\int_0^\infty d\tau\, g(\tau)\, \y(t-\tau),
\label{eq:Delta_def}
\end{equation}
and analize its behavior as $t\to\infty$. We note that for $b>0$, the driver trajectory $\x(t)$ in Eq.~(\ref{eq:SolutionX1Master}) tends to zero at long times (the driver is a damped oscillator). If $\mathbf{\Delta}(t)$ diverges at long times, the components of the driven system $\y(t)$ also diverge and the synchronised solution is unstable. 

It is important to note that the condition $\mathbf{\Delta}(t) \to 0$ as $t \to \infty$ does not, by itself, guarantee the stability of the synchronised state. In particular, both $\mathbf{x}(t)$ and $\mathbf{y}(t)$ may decay to zero without exhibiting synchronisation.

We now analyze the dominant terms in the long-time behaviours of $\x(t)$, $\y(t)$ and $\mathbf{\Delta}(t)$. We start with $\mathbf{\Delta}(t)$. Combining \Eqb{eq:driver_osc}{eq:driven_osc} with \Eq{eq:Delta_def}, we find
\begin{eqnarray}\label{eq:ComputationTimeEvolDeltasGenSum}
 \dot{\Delta}_1(t)&=& \Delta_2(t) - K_1\int_0^\infty \dd{\tau} g(\tau)\Delta_1\mleft(t-\tau\mright), \nonumber \\
 \dot{\Delta}_2(t) &=&
 -\Omega^2\Delta_1(t) - 2b\Delta_2(t) \nonumber \\ 
 &&- K_2\int_0^\infty\dd{\tau}g(\tau)\Delta_2\mleft(t-\tau\mright).
\end{eqnarray}
To solve Eqs.~\eqref{eq:ComputationTimeEvolDeltasGenSum} we perform a Laplace transform. Assuming that $\Delta_i(t<0)=0$ and general values $\Delta_i(t=0)$, we arrive (after some minor algebra) at 
\begin{eqnarray}
 \hat{\Delta}_1(s) &=& \frac{\Delta_1(0) \left[2 b+\hat{g}(s) K_2+s\right]+\Delta_2(0)}{\left[\hat{g}(s) K_1+s\right] \left[2 b+\hat{
g}(s) K_2+s\right]+\Omega ^2}, \nonumber \\
 \hat{\Delta}_2(s) &=& \frac{\hat{g}(s) K_1 \Delta_2(0)+s \Delta_2(0)-\Delta_1(0) \Omega ^2}{\left[\hat{g}(s) K_1+s\right] \left[2 b+\hat{g}(s) K_2+s\right]+\Omega ^2}. \label{eq:DeltaSolutionsLaplace}
\end{eqnarray}

We have written $\hat{f}(s)= \int_0^\infty dt e^{-st}f(t)$ for the Laplace transform of a function $f(t)$. 

To carry out the inverse Laplace transform we use the residue theorem, and obtain $\Delta_i(t) = \sum_{s_k}e^{s_k t}\Res_{s=s_k}\hat{\Delta}_i(s)$, $i=1,2$, where the possible eigenvalues $s_k$ are the poles of $\hat{\Delta}_i(s)$, i.e., the zeroes of the common denominator in Eq.~(\ref{eq:DeltaSolutionsLaplace}).
The dominant behavior of $\Delta_1$ and $\Delta_2$ is determined by the pole with the greatest real part. We write $s_{\max} \pm i\Omega_1$ for this dominant pole, where $s_{\max} = \max_k \Re{s_k}$ and where $\Omega_1$ is the imaginary part of the pole. As a result, the long-term behaviour is of the form $\Delta_i \sim e^{s_{\max}t}\cos\mleft( \Omega_1 t + \phi_i \mright)$ for $i=1,2$, where the $\phi_i$ are phases determined by the initial conditions. If $s_{\max} < 0$ the $\Delta_i$ will approach zero at long times; otherwise they will diverge.

Starting from \Eq{eq:driven_osc} and using the known solution for $\x(t)$ [Eq.~(\ref{eq:SolutionX1Master})], we can carry out a similar Laplace transform analysis of $y_1$ and $y_2$. In this case, we find the poles given by the zeroes of the denominator in Eq.~(\ref{eq:DeltaSolutionsLaplace}), and additionally two poles located at the eigenvalues of the driver system, $-b \pm i\Omega_0$. As a result, the long-time  behaviour of $y_1$ and $y_2$ is of the form $y_i \sim e^{-bt}\cos\mleft(\Omega_0 t + \varphi_i\mright)$ if $s_{\max}<-b$, or $y_i \sim e^{s_{\max}t}\cos\mleft(\Omega_1 t + \psi_i\mright)$ of $s_{\max}>-b$. The $\varphi_i$ and $\psi_i$ are again phases, determined by the initial conditions.

\begin{figure*}[!t]
\centering
 \includegraphics{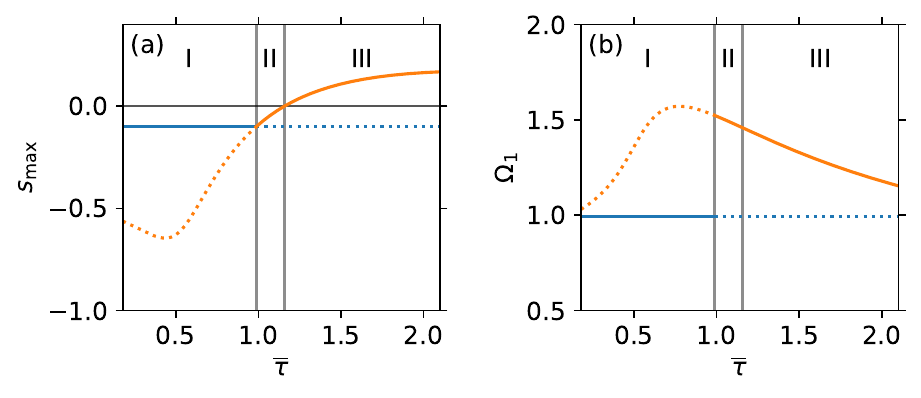}
 \caption{Real part and imaginary parts [panels (a) and (b) respectively] of the leading eigenvalues of the driven system as a function of the average delay $\overline\tau$ for a setup of two harmonic oscillators, with delay distribution $g(\tau) = \frac{1}{2}\left[ \delta\mleft( \tau - \tau_1 \mright) + \delta\mleft( \tau - \tau_2 \mright)\right]$. We set $\Omega = 1$, $b = 0.1$ and $K=0.8$. We also fixed the difference between the two delays $\tau_2 - \tau_1 = 0.2$. The leading eigenvalue is shown by a solid line, and subleading eigenvalue as dotted line. The eigenvalue of the driver system $-b \pm i\Omega_0$ is plotted in blue, while in orange the other leading eigenvalue of the driven system $s_{\max} \pm \Omega_1$ is shown.}
 \label{fig:EigenvaluesSlaveDHO}
\end{figure*}

\medskip

We can distinguish three different regions in parameter space:
\begin{enumerate}[label=\textbf{\Roman*.}]
 \item \textbf{Driven system decays, and shows AS:} In this region we have $s_{\max} < -b$ (and $-b<0$ by definition). All $x_i$ and $y_i$ approach zero as $e^{-bt}$, and $\Delta_i$ as $e^{s_{\max} t}$. The damped oscillations of the driven system have frequency $\Omega_0$ (i.e., the same frequency as the driver).  Given that $s_{\max} < -b$, the coupling terms $\Delta_i$ tend to zero faster than the $x_i$ and $y_i$. This indicates anticipated synchronisation at long times. 
 \item \textbf{Driven system decays, but no AS:} Here, $-b < s_{\max} < 0$. The $x_i$ and $y_i$ again tend to zero at long times, and as a consequence the $\Delta_i$ also tend to zero. However, the decay of the $\Delta_i$ is slower than that of the $x_i$. Further, the $y_i$ oscillate with a frequency $\Omega_1$ different to the frequency of the driver system. Thus, there is no synchronisation. 
 \item \textbf{Driven system diverges:} This region is characterised by $s_{\max} > 0$, as a result both $\Delta_i$ and $y_i$ diverge at long times. There is no synchronisation.
\end{enumerate}

\subsection{Example: System with two delay channels}
So far, our analysis is very general and independent of the specific form $g(\tau)$ of the delay distribution. We now apply these results to the functional form
\begin{equation}\label{eq:g2taus}
g(\tau) = \frac{1}{2}\left[ \delta\mleft( \tau - \tau_1 \mright) + \delta\mleft( \tau - \tau_2 \mright)\right].
\end{equation}
This describes a setup with two delay lines between the driver and the driven systems, with delay times $\tau_1$ and $\tau_2$ respectively. The two delays carry equal weights. 

We define $\Delta \tau=\tau_2-\tau_1$. Using this in \Eqb{eq:beta}{eq:tstar}, we find

\begin{equation}\label{eq:taubar_2delays}
\tau^\ast=\overline\tau+\Omega_0^{-1}\arctan\left[\tan(\Omega_0\frac{\Delta \tau}{2})\tanh(b\frac{\Delta \tau}{2})\right].
\end{equation}
Here $\overline{\tau}=(\tau_1+\tau_2)/2$ is the average value of the delays. We also find
\begin{equation}
\beta= \frac{\sqrt{2}\exp{b\Omega_0^{-1}\arctan\left[\tan(\Omega_0\frac{\Delta \tau}{2})\tanh(b\frac{\Delta \tau}{2})\right]}}{\sqrt{\cos(\Omega_0\Delta\tau)+\cosh(b\Delta\tau)}}.
\end{equation}

One can expand the anticipation time and the amplification factor in terms of $\Delta$ as follows
\begin{eqnarray}\label{eq:TaylorExpParametersTwoDelays}
 \tau^\ast &=& \bar{\tau} + \frac{b}{4}(\Delta\tau)^2 + {\cal O}[(\Delta\tau)^4], \\
 \beta &=& 1 + \frac{\Omega_0^2 + b^2}{8}(\Delta\tau)^2 + {\cal O}[(\Delta\tau)^4].
\end{eqnarray}

\begin{figure*}[!t]
 \centering
 \includegraphics{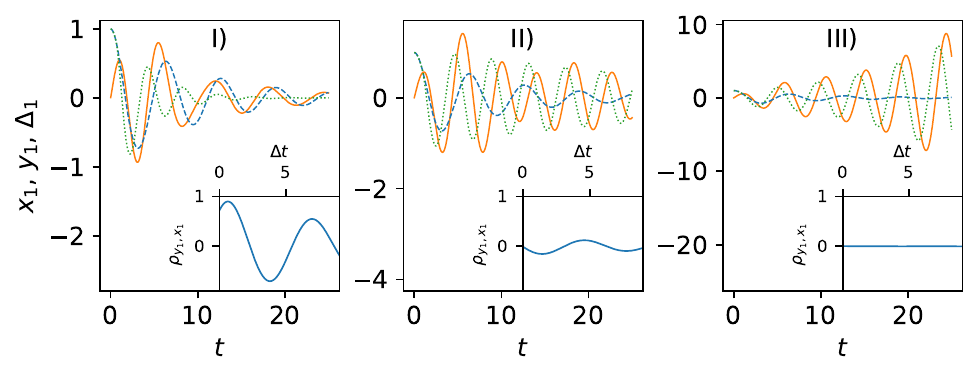}
 \caption{Sample trajectories from numerical integration of the system in \Eqb{eq:driver_osc}{eq:driven_osc} with the delay distribution of \Eq{eq:g2taus}, and initial conditions $y_i(t\le 0)=0$, $x_1(0)=1,x_2(0)=0$. Parameters were chosen to exemplify the whole phase space. I) The driven system shows anticipated synchronisation of the driver: $\tau_1 = 0.7$, $\tau_2 = 0.9$, II) the driven system is not synchronised with the driver but it decays to zero: $\tau_1 = 1$, $\tau_2 = 1.2$, and III) the driven system diverges: $\tau_1 = 1.3$, $\tau_2 = 1.5$. The other parameters are the same for all the panels: $\Omega = 1$, $b = 0.1$, $K_1 = 0.8$ and $K_2 = 0$. The solid orange lines correspond with the driven system's variable ($y_1$), the dashed blue lines with the driver system's variable ($x_1$), and the dotted green lines with the coupling mismatch $\Delta_1$, see Eq.~\eqref{eq:Delta_def}. An inset is added to each of the panels to show the correlation function between the $x_1$ and the $y_1$ variables. }
 \label{fig:TrajectoryDampedOsc}
\end{figure*}

The leading-order corrections in $\Delta \tau$ indicate an increase of the time shift and of the amplitude of the driven system (at fixed driver amplitude) for small but non-zero $\Delta\tau$. A similar analysis can be carried out if the two delays $\tau_1$ and $\tau_2$ carry different weights. In that case the time shift will be the weighted average of the two delays, plus corrections. The same system can also be studied for uniformly distributed delay, some details are given in \ref{app:harm_osc_uniform_delay}. 

We now illustrate these results. In order to reduce the number of available parameters, we set $\Omega=1$, $b=0.1$, $\Delta\tau=0.2$. We also set $K_2=0$; this represents a minimal coupling between the driven system and the driver. We then study the different dynamical regimes as a function of the average delay $\overline\tau$ and coupling constant $K_1$, which we write as $K$ for simplicity. 

In Figure~\ref{fig:EigenvaluesSlaveDHO} we plot the real part $s_{\max}$, and the imaginary part $\Omega_1$, of the leading eigenvalue as a function of $\overline\tau$ for fixed $K=0.8$. Details of the numerical calculation of the eigenvalues are given in \ref{app:roots}. We can identify the three scenarios described above. For sufficiently small $\overline\tau$, we have $s_{\max}<-b$ (this is the AS regime I). Increasing $\overline\tau$ leads to a regime in which $-b<s_{\max}<0$ such that, while the coupling mismatch $\mathbf{\Delta}$ between the two oscillators tends to zero, there is no AS (regime II). Increasing $\overline\tau$ even further we find $s_{\max}>0$ so that the driven system diverges (regime III).

The corresponding trajectories in the three regimes are plotted in Fig.~\ref{fig:TrajectoryDampedOsc}. In panel (I) a trajectory with AS is shown. The driven system also decays with time, but the trajectory of the driven system anticipates that of the driver. The quantities $\Delta_i$ decay to zero faster than the signals $x_i$ and $y_i$. We note that, even though the coupling between the driver and driven systems only acts on the components $y_1$, the second component $y_2$ also anticipates $x_2$ in this regime. In panel (II) the $x_i$ and $y_i$ are no longer synchronized with one another. The components $y_i$ of the driven system decay more slowly than the driver components $x_i$. The quantities $\Delta_i$ decay at the same rate as $y_i$. Finally, in panel~(III) the $y_i$ and $\Delta_i$ diverge at long times.

\begin{figure}
 \centering
 \includegraphics[width=0.4\textwidth]{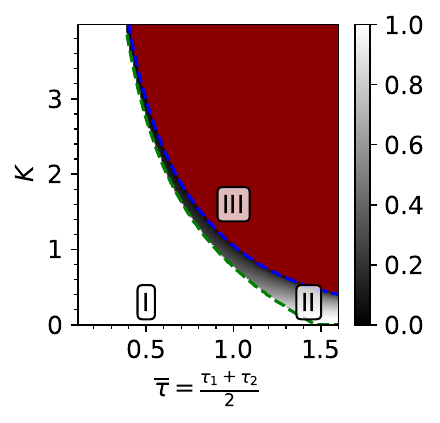}
 \caption{Stability diagram for the setup of two damped harmonic oscillators \Eqb{eq:driver_osc}{eq:driven_osc} with the delay distribution of \Eq{eq:g2taus}. Below the green dashed line the theory predicts the presence of AS (region I). Between the two dashed lines the analytical calculation shows decaying behaviour of the driver and the driven system, but no AS (region II). Above the blue dashed line finally, we expect the driven system to diverge (region III). The background colour shading in regions I and II indicates the value of $\left[ \max_{\Delta t}\mleft\{\rho_{x_1, y_1}(\Delta t)\mright\} + \max_{\Delta t}\mleft\{\rho_{x_2, y_2}(\Delta t)\mright\} \right]/2$, obtained from numerical integration of \Eqb{eq:driver_osc}{eq:driven_osc}. Combinations of $\overline\tau$ and $K$ are shown in red when the height of the maxima of the driven system exceeds that of the driver by a factor of at least $100$ at the end of the simulation $T_{\mathrm{sim}} = 6000$. We use this as an indicator of divergence in the driven system. Model parameters are $\Omega = 1$ and $b = 0.1$. We fix $\tau_2-\tau_1 = 0.2$.}
 \label{fig:StabilityDampOsc}
\end{figure}
By repeating this analysis for different values of $K$ we obtain the stability diagram in Fig.~\ref{fig:StabilityDampOsc}. The dashed green line separates regime I (to the left of the line) from regime II, while the dashed blue line separates regimes II and III. The background colour in the figure indicates the average correlation function $\left(\left[ \max_{\Delta t}\mleft\{\rho_{x_1, y_1}(\Delta t)\mright\} + \max_{\Delta t}\mleft\{\rho_{x_2, y_2}(\Delta t)\mright\} \right]/2\right)$, computed from a direct numerical integration of \Eqb{eq:driver_osc}{eq:driven_osc}.

Broadly, for sufficiently small average delay $\overline\tau$ or coupling strength $K$ we find decaying trajectories with AS (region I, coloured white in the diagram). For large $\overline\tau$ or $K$ the driven system diverges (region III, red). There is an intermediate range of parameters (region II, grey scale colouring in the figure indicates the maximum value of the cross correlation between driver and driven systems) in which the driver and the driven system both decay in time, but where there is no AS. As detailed in the figure caption results from a direct numerical integration of \Eqb{eq:driver_osc}{eq:driven_osc} confirm the predictions of the theory.

\begin{figure}
 \centering
 \includegraphics[width=0.4\textwidth]{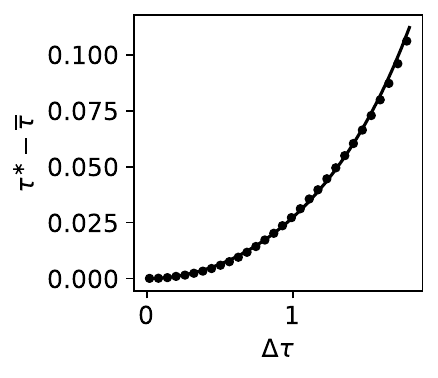}
 \caption{Anticipation time ($\tau^\ast$) from the theory and the numerical integration of the system of the system in \Eqb{eq:driver_osc}{eq:driven_osc} with the delay distribution of \Eq{eq:g2taus}. The results are shown as a function of $\Delta\tau = \tau_2 - \tau_1$. The parameters are: $\overline{\tau} = 0.9$, $\Omega = 1$, $b = 0.1$, $K_1 = 0.8$ and $K_2 = 0$. The lines are the results from the theory, Eq.~\eqref{eq:taubar_2delays}, and the points are obtained from numerical integration of the original system. Note that $\Delta\tau=1.8$ is the maximum possible value given $\overline\tau=0.9$.}
 \label{fig:TimeshiftVsDeltaTauDampedOsc}
\end{figure}

Numerically, we extract the anticipation time $\tau^\ast$ from the location of the the maximum of the correlation function $\rho_{y_1, x_1}(\Delta t)$. In Fig.~\ref{fig:TimeshiftVsDeltaTauDampedOsc} we compare this numerical value with the theoretical result of Eq.~(\ref{eq:taubar_2delays}), and find agreement. As indicated by Eq.~(\ref{eq:taubar_2delays}) the difference $\tau^\ast-\overline\tau$ depends only the delay difference $\Delta\tau=\tau_2-\tau_1$ but not on $\overline{\tau}$. When the delay difference is zero the system reduces to the single delay case, $\tau_1 = \tau_2 = \overline{\tau} = \tau^\ast$ (left end of the figure).

We now compare the results obtained for two delay lines, Eq.~\eqref{eq:g2taus}, with those of a single delay, $g(\tau)=\delta(\tau-\tau_s)$. Our goal is to assess whether the presence of two delay lines can enlarge the region of stability of anticipated synchronisation. To this end, we fix the resulting anticipation time $\tau^\ast$ and compare the corresponding stability diagrams in the parameter space spanned by $\tau^\ast$ and $K$. Details of the numerical procedure used to determine the stability of the AS solution can be found in \ref{app:roots}.

In the single-delay case, fixing $\tau^\ast$ requires setting $\tau_s=\tau^\ast$. In contrast, for the two-delay configuration there exists a continuous range of pairs $(\tau_1,\tau_2)$ that, according to Eq.~\eqref{eq:tstar}, yield the given anticipation time $\tau^\ast$. When $\tau_1=\tau_2=\tau^\ast$, the two-delay configuration reduces to the single-delay case. To illustrate the effect of genuinely distinct delay lines, we compare this single-delay case, $\tau_1=\tau^\ast$, with two representative situations: the extreme case $\tau_1=0$ and an intermediate case $\tau_1=\tau^\ast/2$ (and with $\tau_2$ suitably adjusted to produce the given value of $\tau^\ast$).
\begin{figure}
 \centering
 \includegraphics[width=0.4\textwidth]{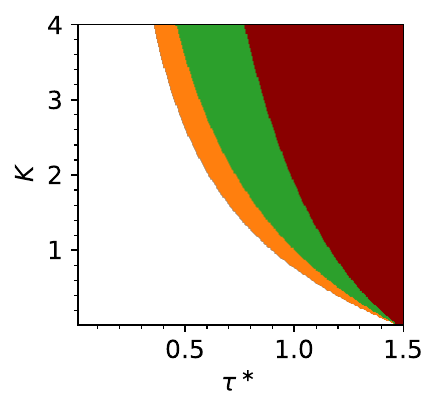}
 \caption{Comparison of the occurrence of AS in the damped oscillator system with a single delay ($\tau_1 = \tau^\ast = \tau_2$) and two models with two delay channels (one of these setup has $\tau_1 = \tau^\ast/2$ and the other $\tau_1 = 0$). See the main text for details. In the white region all three setups show AS, in the orange region the setup with a single delay ($\tau_1 = \tau^\ast$) no longer shows AS, but the other two setups continue to have AS. In the green region only the setup with $\tau_1 = 0$ exhibits AS, and in the red region no setup has AS. Model parameters are $\Omega = 1$ and $b = 0.1$. In the model with two delays we set $\tau_1$ as indicated; the value of $\tau_2$ is determined to produce the desired value of $\tau^\ast$. }
 
 \label{fig:ComparisonTwoDelaysSingle}
\end{figure}

The results of this comparison are shown in Fig.~\ref{fig:ComparisonTwoDelaysSingle}. The white region corresponds to parameters for which anticipated synchronization is observed in all three models (single delay $\tau_1 = \tau^\ast$; two delays with $\tau_1 = \tau^\ast/2$, and two-delay system with $\tau_1 = 0$). The orange region indicates parameter values for which anticipated synchronization no longer exist in the single-delay case ($\tau_1 = \tau^\ast$), but where it continues to exist in the other two models. In the green region only the model with $\tau_1=0$ exhibits AS, while the red region corresponds to the absence of anticipated synchronization in all three models. These results show systems with distributed delay can support anticipated synchronization with larger anticipation times $\tau^\ast$ than setups with only a single delay.

\begin{figure}
 \centering
 \includegraphics[width=0.4\textwidth]{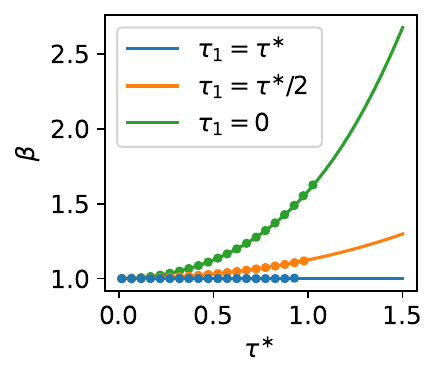}
 \caption{Amplification factor $\beta$ from the theory and from numerical integration of the system in \Eqb{eq:driver_osc}{eq:driven_osc} with the delay distribution of \Eq{eq:g2taus}. The only parameters that are changed are the two delays of the distribution ($\tau_1$ and $\tau_2$), all the other parameters are fixed ($\Omega = 1$, $b = 0.1$, $K_1 = 0.8$). and $K_2 = 0$. The lines are the results from the theory and the points are obtained from numerical integration. To generate the data, we fix $\tau_1$, and $\tau_2$ is then chosen to generate the desired anticipation time $\tau^\ast$. The numerical results can only be plotted when the solution is stable.}
 \label{fig:BetaTauAstDampedOsc}
\end{figure}

Finally, we examine how the amplification factor $\beta$ depends on the anticipation time $\tau^\ast$. For the single-delay configuration $\tau_1=\tau^\ast$ we have $\beta(\tau^\ast)=1$ independently of $\tau^\ast$. In Fig.~\ref{fig:BetaTauAstDampedOsc} we compare this with the amplification factor for the two two-delay configurations  above ($\tau_1=\tau^\ast/2$ and $\tau_1=0$, respectively). As seen in the figure, the amplification factor increases with $\tau^\ast$ and, for a given $\tau^\ast$, reaches its largest values in the case $\tau_1=0$.

\subsection{Two-dimensional damped oscillator}
\label{subsec:coupled_oscillators}

We now extend the analysis to a system in which the driver and the driven units each consist of two coupled damped harmonic oscillators. This configuration introduces multiple intrinsic time scales through distinct normal modes and therefore provides a natural setting to assess the robustness of anticipated synchronisation beyond the single-mode case considered so far.

The driver system is described by
\begin{align}
\dot{x}_1 &= x_2, \nonumber\\
\dot{x}_2 &= -\Omega^2 x_1 - k (x_1 - x_3) - 2b x_2, \nonumber\\
\dot{x}_3 &= x_4, \nonumber\\
\dot{x}_4 &= -\Omega^2 x_3 + k (x_1 - x_3) - 2b x_4,
\label{eq:driver_coupled}
\end{align}
where $x_1, x_2$, and $x_3, x_4$, respectively, denote the variables of the first and second oscillators. The quantity $k$ is the coupling strength between the oscillators (we stress that this is a coupling {\em within} the driver, not between the driver and the driven system). As before, $\Omega$ is the natural frequency, and $b$ the damping coefficient. This system has two normal modes with frequencies
\begin{equation}
\omega_{\mathrm{s}} = \sqrt{\Omega^2 - b^2}, \qquad
\omega_{\mathrm{a}} = \sqrt{\Omega^2 + 2k - b^2},
\end{equation}
corresponding to in-phase and anti-phase oscillations of the two oscillators in the driver unit.

The driven system is an identical copy of the driver, subject to a distributed delay coupling acting on the first oscillator only,
\begin{align}
\dot{y}_1 &= y_2
+ K \left[
x_1(t) - \int_0^\infty d\tau\, g(\tau)\, y_1(t-\tau)
\right], \nonumber\\
\dot{y}_2 &= -\Omega^2 y_1 - k (y_1 - y_3) - 2b y_2, \nonumber\\
\dot{y}_3 &= y_4, \nonumber\\
\dot{y}_4 &= -\Omega^2 y_3 + k (y_1 - y_3) - 2b y_4 .
\label{eq:driven_coupled}
\end{align}
As before, the constant $K$ describes the strength of the coupling between the driver and the driven units, and $g(\tau)$ is a normalized delay distribution. The driver evolves autonomously. 

Due to the linearity of the system, the solution for the driver component $x_1(t)$ can be written as a superposition of the two normal modes,
\begin{eqnarray}
 x_1(t) &=&
 A_{\mathrm{s}} e^{-bt} \cos(\omega_{\mathrm{s}} t + \phi_{\mathrm{s}})
 + A_{\mathrm{a}} e^{-bt} \cos(\omega_{\mathrm{a}} t + \phi_{\mathrm{a}}) \nonumber \\
 &\equiv& x_{1,\mathrm{s}}(t) + x_{1,\mathrm{a}}(t),
 \label{eq:x1_modes}
\end{eqnarray}
with constants $A_\mathrm{s}$, $A_\mathrm{a}$, $\phi_{\mathrm{s}}$ and $\phi_{\mathrm{a}}$ determined by initial conditions. For simplicity we have defined $x_{1,\mathrm{s}}(t) \equiv A_{\mathrm{s}}e^{-bt}\cos(\omega_{\mathrm{s}}t + \phi_{\mathrm{a}})$, and similarly for $x_{1\mathrm{a}}$. 

Proceeding as in Sec.~\ref{sec:single_oscillator}, one can again impose the condition that the delayed coupling term vanishes. Owing to the superposition principle, each normal mode can be treated independently. As a result, the  component $y_1(t)$ in the driven system takes the form
\begin{equation}\label{eq:y1_modes}
y_1(t) =
\beta(\omega_{\mathrm{s}}) x_{1,\mathrm{s}}\!\left[t+\tau^\ast(\omega_{\mathrm{s}})\right] + \beta(\omega_{\mathrm{a}}) x_{1,\mathrm{a}}\!\left[t+\tau^\ast(\omega_{\mathrm{a}})\right],
\end{equation}
where the amplification factors $\beta(\omega_\mathrm{s})$ and $\beta(\omega_\mathrm{a})$, and the effective anticipation times $\tau^\ast(\omega_\mathrm{s})$ and $\tau^\ast(\omega_\mathrm{a})$ are given by \Eqb{eq:beta}{eq:tstar}, with $\Omega_0$ replaced by $\omega_{\mathrm{s}}$ or $\omega_{\mathrm{a}}$ respectively. An analogous expression holds for the other components $y_2(t)$, $y_3(t)$, and $y_4(t)$.

Equation~\eqref{eq:y1_modes} shows that, in contrast to the single-oscillator case, the driven trajectory cannot be obtained from the driver by a single global time shift and amplitude rescaling. Each normal mode anticipates the driver by a different amount and is amplified by a different factor. As a consequence, exact anticipated synchronisation no longer occurs in this system.
\begin{figure}[h]
 \centering \includegraphics[width=0.4\textwidth]{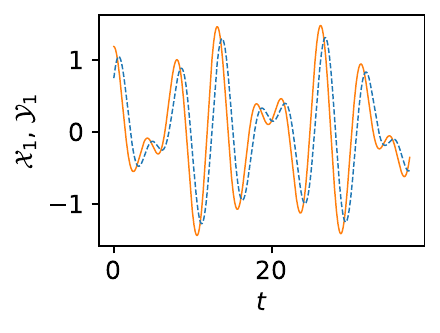}
 \caption{Example of re-scaled trajectories of the driver and driven systems for a setup formed by two units of two coupled damped harmonic oscillators each. The quantities shown are ${\cal X}_1(t)\equiv e^{bt}x_1(t)$ in a dashed blue line and ${\cal Y}_1(t)\equiv e^{bt}y_1(t)$ in a solid orange line (see text for further explanation). Initial conditions $x_2(0)=x_3(0)=1,x_1(0)=x_4(0)=0$, $y_i(t\le0)=0$. Parameters are $\Omega = 1$, $k = 0.5$, and $b = 0.1$. In the coupling term two equally weighted delays were used: $\tau_1 = 0.2$ and $\tau_2 = 1.0$. The coupling strength is $K = 1$.}
 \label{fig:ExampleTrajectoryTwoCDHO}
\end{figure}

\begin{figure*}[!t]
 \centering
 \includegraphics[width=0.8\textwidth]{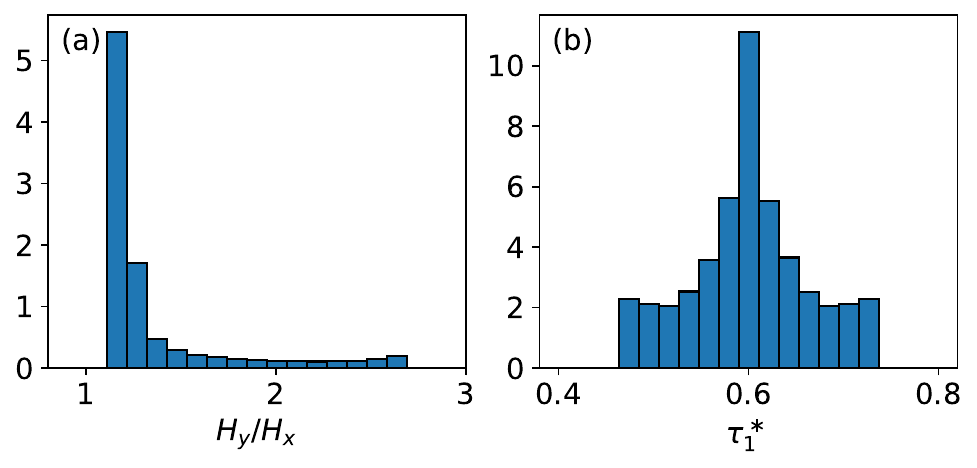}
 \caption{Ratio of amplitudes and anticipation time distributions for two coupled damped harmonic oscillators with parameters $\Omega = 1$, $k = 0.5$, and $b = 0.1$. 
 Panel (a) shows the histogram of the ratio $H_{y}/H_{x}$, where $H_{x}$ denotes the absolute difference between two consecutive extrema (one minimum and the subsequent maximum) of the driver system, and $H_{y}$ is the absolute difference between the corresponding, anticipated, extrema of the driven system. Panel (b) shows a histogram of the anticipation time $\tau_1^\ast$ defined as the time lag between an extremum (maximum or minimum) of the driven system and the subsequent extremum of the driver system. Data is for two equally weighted delays $\tau_1 = 0.2$ and $\tau_2 = 1.0$, with coupling strength $K = 1$. These plots are constructed from the variables $x_1$ and $y_1$, but similar results can be found for the other variables.}
	\label{fig:AmpsAndShiftsTwoCDHO}
\end{figure*}

Nevertheless, numerical integration of \Eqb{eq:driver_coupled}{eq:driven_coupled} reveals behaviour that closely resembles anticipated synchronization. As illustrated in Fig.~\ref{fig:ExampleTrajectoryTwoCDHO}, the driven system qualitatively reproduces the main features of the driver trajectory, with extrema systematically occurring earlier in time. In the figure we plot the rescaled variables
\begin{equation}
{\cal X}_1(t) = e^{bt} x_1(t), \qquad
{\cal Y}_1(t) = e^{bt} y_1(t).
\end{equation}
This removes the exponential decay resulting from the damping, and facilitates a qualitative visual comparison of the driver and driven signals.

A closer inspection shows that both the effective time shift and the amplification vary over time. We characterize these effects by comparing successive minima and maxima of $x_1(t)$ and $y_1(t)$, from which we construct an amplification measure as explained in the caption of Fig.~\ref{fig:AmpsAndShiftsTwoCDHO}. In the figure we show the distribution of the instantaneous amplification measure and time shift.  This confirms consistent amplification and anticipation. The mean estimated anticipation time is found to be close to the mean delay of the distribution, similar to the analytical predictions for the simpler system in Sec.~\ref{sec:single_oscillator}.

In summary, while the presence of multiple intrinsic frequencies prevents exact anticipated synchronization, systems of coupled harmonic oscillators with distributed delay coupling can still exhibit robust, approximate anticipation, accompanied by systematic amplification of the driven response.

\section{Chaotic systems}
\label{sec:chaotic}

We now turn to nonlinear chaotic systems with distributed delay coupling. Analytical progress is generally not possible here, and we therefore rely on numerical integration.

\subsection{R\"ossler system}
\begin{figure*}[!t]
 \centering
 \includegraphics{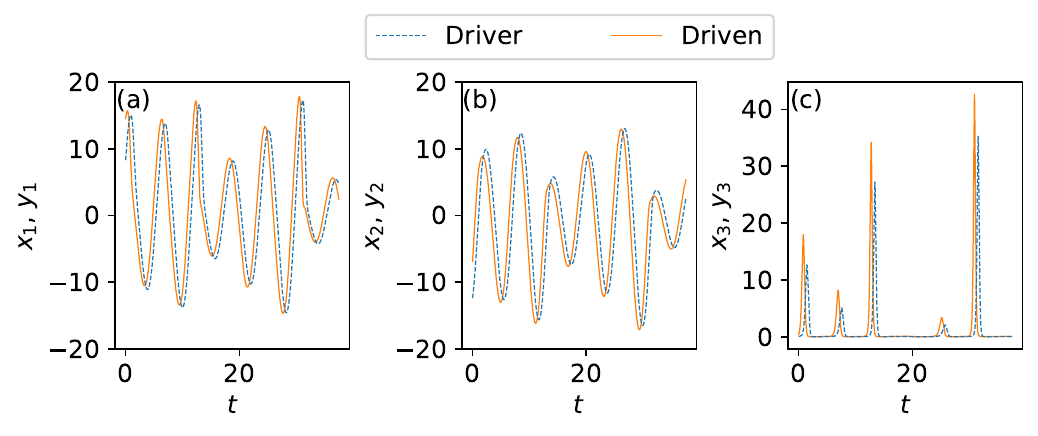}
 \caption{Trajectory for the R\"ossler system with a uniformly distributed delay in the interval $\left[ 0.1,\ 1.1 \right]$. The values of the parameters are $a = 0.15$, $b = 0.2$, and $c = 10$. The coupling strength is $K = 1$. The time $t=0$ is chosen arbitrarily in the stationary state. The data is from numerical integration of \Eqb{eq:RosslerDriver}{eq:RosslerDriven}.}
 \label{fig:TrajUnifDistRossler}
\end{figure*}
\begin{figure*}[!t]
	\centering
	\includegraphics[width=0.8\textwidth]{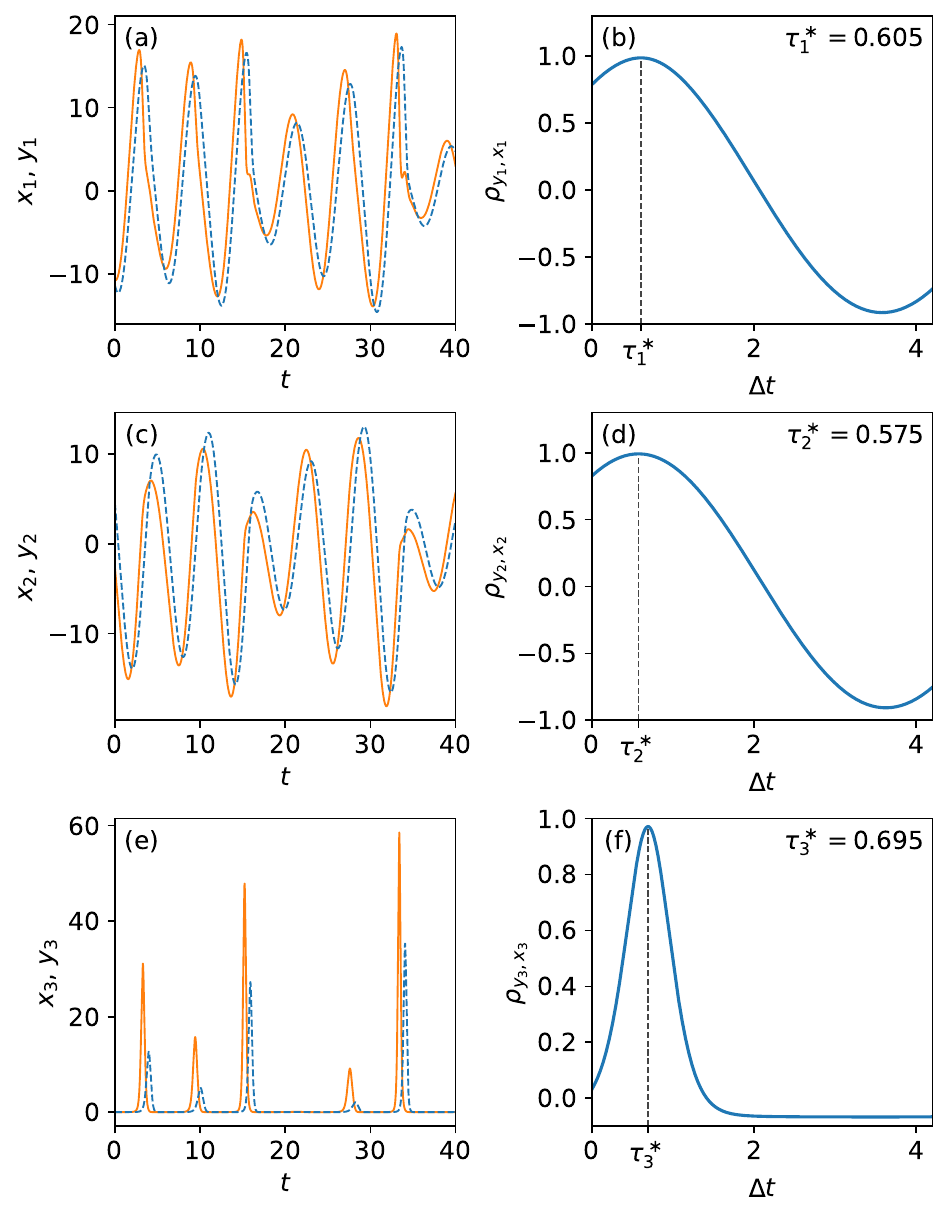}
	\caption{Trajectory for the R\"ossler system with two equally weighted delays: $\tau_1 = 0.1$ and $\tau_2 = 1.1$.
 The value of the parameters are $a = 0.15$, $b = 0.2$, and $c = 10$. The coupling strength is $K = 1$. The time $t=0$ is chosen arbitrarily in the stationary state. The right-hand panels show the cross-correlation functions between driver and driven systems for each of the three components of the dynamics. The trajectory of the driver variable is plotted in dashed blue lines and the trajectory of the driven system is in solid orange lines.}
	\label{fig:TrajRossler2Delays}
\end{figure*}

As a first example we consider the R\"ossler system, for which anticipated synchronization with a single delay has been extensively studied~\cite{Voss_2000, Voss_2001a, Pyragas_2008}. The autonomous driver dynamics is given by
\begin{align}
\dot{x}_1 &= -x_2 - x_3, \nonumber\\
\dot{x}_2 &= x_1 + a x_2, \nonumber\\
\dot{x}_3 &= b + x_3(x_1 - c),
\label{eq:RosslerDriver}
\end{align}
with parameters $a$, $b$, and $c$. The driven system follows identical intrinsic dynamics, with additional coupling to the driver system involving distributed delay. We have
\begin{align}
\dot{y}_1 &= -y_2 - y_3
+ K\!\left[
x_1(t) - \int_0^\infty d\tau\, g(\tau)\, y_1(t-\tau)
\right], \nonumber\\
\dot{y}_2 &= y_1 + a y_2, \nonumber\\
\dot{y}_3 &= b + y_3(y_1 - c),
\label{eq:RosslerDriven}
\end{align}
where $K$ denotes the coupling strength.

An example of a trajectory for uniformly distributed delay in the interval $[0.1,1.1]$ is shown in Fig.~\ref{fig:TrajUnifDistRossler}. While we do not observe exact anticipated synchronization, the time series of the driver and driven systems are similar to one another, modulo an appropriate time shift. In particular, the extrema of the driven variables consistently occur earlier than those of the driver, indicating anticipated behaviour. The effective anticipation time for each component, extracted from the position of the maximum of the cross-correlation function, is close to the mean of the delay distribution $\overline\tau=0.6$ ($\tau^\ast_1 = 0.575$, $\tau^\ast_2 = \tau^\ast_3 = 0.580$). 

In addition to anticipation, Fig.~\ref{fig:TrajUnifDistRossler} also reveals amplification effects. In particular, the maxima of the driven variable $y_3$ exceed those of the corresponding driver variable $x_3$ over extended time intervals. This indicates that, as in the linear systems studied earlier, distributed delay coupling can lead to an amplification of the driven response relative to the driver.

A second example is shown in Fig.~\ref{fig:TrajRossler2Delays} for a delay distribution consisting of two equally weighted delays [\Eq{eq:g2taus}]. Again, approximate anticipated synchronization is clearly visible. For the pairs $(x_1,y_1)$ and $(x_2,y_2)$ [panels (a) and (c), respectively], the sequence of maxima and minima in the driven system closely mirrors that of the driver, but occurs earlier in time. The third variables $(x_3,y_3)$ [panel (e)] exhibit more spike-like behavior, but again the driven system anticipates the driver and shows systematically larger excursions.

The anticipation is confirmed quantitatively by the cross-correlation functions shown in panels (b), (d), and (f) of Fig.~\ref{fig:TrajRossler2Delays}. The maxima of the correlation functions occur at positive time shifts, with values $0.605$, $0.575$, and $0.695$ for the three components respectively, while the mean delay in this setup is $0.6$. This confirms that, although the anticipation times vary between components, they remains close to the average delay.

\begin{figure*}[!t]
	\centering
 \includegraphics[width=0.7\textwidth]{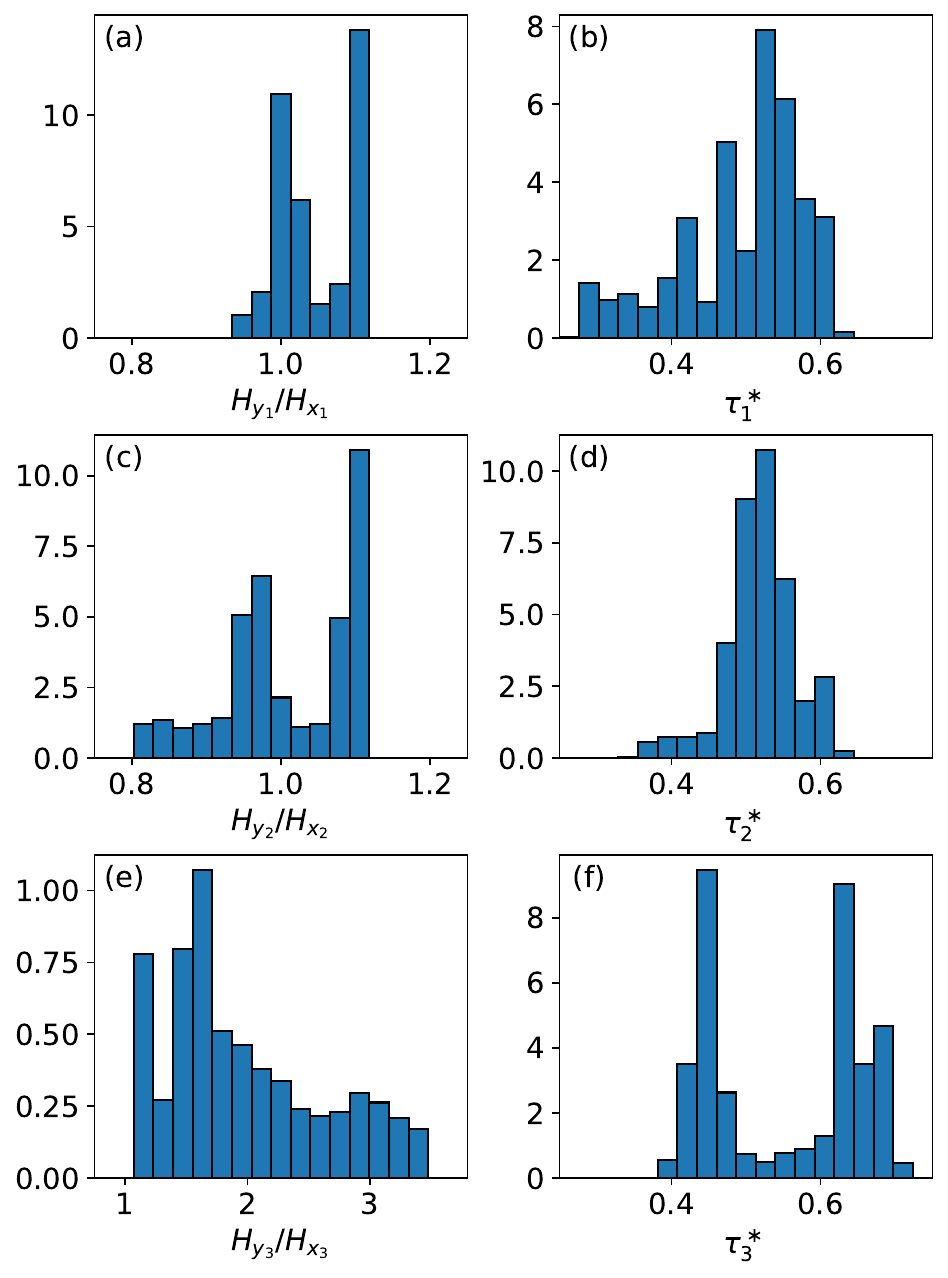}
	\caption{Ratio of amplitudes $H_y/H_x$ an anticipation times $\tau^\ast$ for the Rössler system given in \Eqb{eq:RosslerDriver}{eq:RosslerDriven} with two equally weighted delays $\tau_1 = 0.1$ and $\tau_2 = 1.0$. The quantities plotted here have already been defined in Fig.~\ref{fig:AmpsAndShiftsTwoCDHO}. The values of the parameters are $a = 0.15$, $b = 0.2$, and $c = 10$. The coupling strength is $K = 1$. Note that the scale of the horizontal axis in panel (e) is different from that in panels (a) and (c).}
	\label{fig:DistributionAmpShiftRossler2Delays}
\end{figure*}

To characterize amplification and anticipation more systematically, we compute instantaneous amplification factors and time shifts between successive extrema, as introduced earlier for linear systems. This analysis is carried out separately for each of the three components of the R\"ossler system. We show the resulting distributions in Fig.~\ref{fig:DistributionAmpShiftRossler2Delays}. In contrast to the harmonic oscillator case, the amplification measure $H_y/H_x$ now takes values above and below one, indicating that amplification is not a consistent feature at all times. However, values of $H_y/H_x$ greater than one are more frequent than values below one [see panels (a), (c) and (e)], indicating a net tendency towards amplification. Similarly, the instantaneous time shifts in panels (b), (d) and (f) are always positive, confirming that the driven system anticipates the driver.

We obtain qualitatively similar distributions of amplification and anticipation time for uniformly distributed delay (results not shown). We find that, for the same mean delay and compared to the case of two equally weighted delays, a uniform delay distribution leads to a similar mean amplification factor for variables $x_1$ and $y_1$ and a reduced mean amplification factor for variable $x_3$. On the other hand, the variance of the distribution of the amplification factor is reduced for all three variables.

\begin{figure*}
 \includegraphics{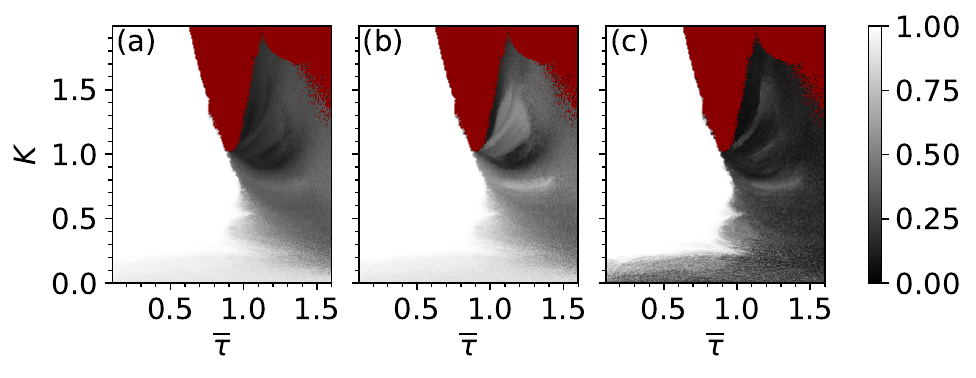}
 \caption{Height at the maximum of the cross-correlation function for the R\"ossler system given in \Eqb{eq:RosslerDriver}{eq:RosslerDriven} with two equally weighted delays. Approximate anticipated synchronisation is indicated by a maximum value of the cross-correlation functions close to one. The results obtained for the different variables are shown in the different panels: $x_1$ and $y_1$ in panel (a), $x_2$, $y_2$ in panel (b), and $x_3$ and $y_3$ in panel (c). Model parameters are $a = 0.15$, $b = 0.2$, and $c = 10$. The difference between the two delays is fixed at $\Delta\tau = 0.2$. In the red regions in the $(K,\overline{\tau})$ plane the driven system diverges.}
 \label{fig:StabRossler2Delays}
\end{figure*}

Finally, we identify the region of parameter space in which approximate anticipated synchronisation occurs in the Rössler system with distributed delays. Focusing on the case of two equally weighted delays, we fix the delay difference $\Delta\tau=\tau_2-\tau_1$ and vary the mean delay $\overline{\tau}=(\tau_1+\tau_2)/2$ together with the coupling strength $K$. 

The heatmap plots in Fig.~\ref{fig:StabRossler2Delays} show the maximum value of the cross-correlation function for the three different variables. Approximate anticipated synchronisation is observed when the maximum correlation $\rho_{y_n,x_n}(\Delta t)$ is close to unity and when the maximum occurs for positive $\Delta t$. As indicated by the white regions in the three panels, approximate anticipated synchronisation is predominantly found when the mean delay and the coupling strength are not too large.

We now briefly discuss the anticipation time $\tau^\ast_i$ for the three components ($i=1,2,3$). In general, we find this to be close to the average $\overline\tau$. More precisely, we quantify the overall relative deviation between $\tau_i^\ast$ and $\overline \tau$ using the symmetric mean absolute percentage error, defined as $\mathrm{SMAPE} = \left\langle 2\frac{\abs{\tau_i^\ast - \overline{\tau}}}{\abs{\tau_i^\ast} + \abs{\overline{\tau}}}\right\rangle$, where the average is taken over all points in which the maximum correlation is greater than a threshold of $0.9$. In all three panels of Fig.~\ref{fig:StabRossler2Delays} this quantity is found to be smaller than $7.5\%$.

\subsection{Lorenz system}
We now turn to a setup of two coupled Lorenz systems. The driver dynamics is given by
\begin{align}
\dot{x}_1 &= a(x_2 - x_1), \nonumber\\
\dot{x}_2 &= x_1(b - x_3) - x_2, \nonumber\\
\dot{x}_3 &= x_1 x_2 - c x_3,
\label{eq:Lorenz_driver}
\end{align}
where $a, b$ and $c$ are parameters. The driven system evolves according to
\begin{align}
\dot{y}_1 &= a(y_2 - y_1)
+ K\!\left[
x_1(t) - \int_0^\infty d\tau\, g(\tau)\, y_1(t-\tau)
\right], \nonumber\\
\dot{y}_2 &= y_1(b - y_3) - y_2, \nonumber\\
\dot{y}_3 &= y_1 y_2 - c y_3.
\label{eq:Lorenz_driven}
\end{align}

\begin{figure*}
	\centering
		\includegraphics[width=0.8\textwidth]{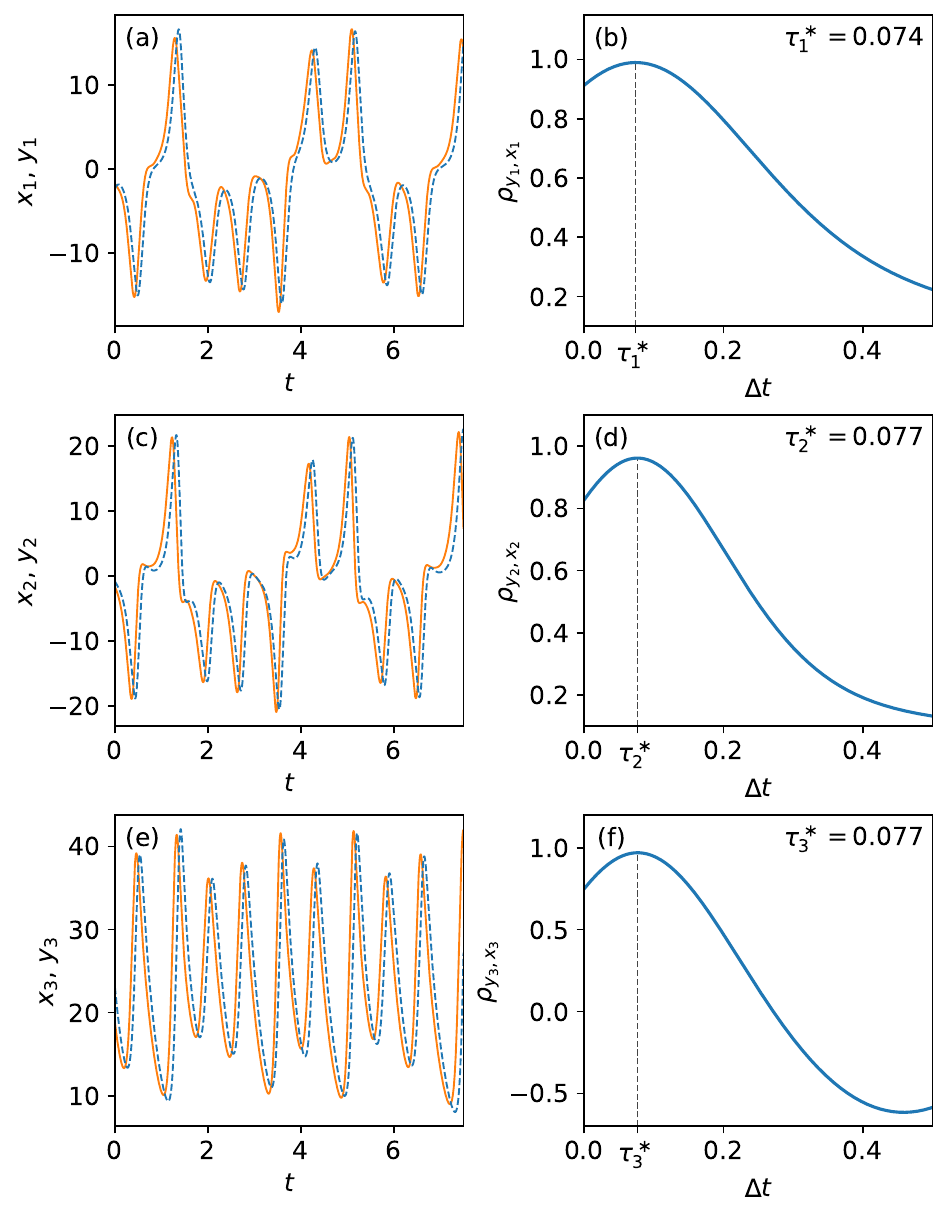}
	\caption{Example of a trajectory for a setup of two coupled Lorenz systems, with uniformly distributed delay in the interval from $\tau_1 = 0.03$ to $\tau_2 = 0.13$. The right-hand panels show the cross-correlation functions between driver and driven systems for each of the three components of the dynamics. Remaining model parameters are $a = 10$, $b = 28$, $c = 8/3$, and $K = 20$. The trajectory of the driver variable is plotted in dashed blue lines and the trajectory of the driven system is in solid orange lines. The time $t=0$ is chosen arbitrarily in the stationary state. }
	\label{fig:TrajLorenz2Delays}
\end{figure*}

Representative trajectories for uniformly distributed delay are shown in Fig.~\ref{fig:TrajLorenz2Delays}. As in the R\"ossler system, the trajectory of the driven system closely follows that of the driver, modulo a time shift. The extrema of the variables in the driven system occur earlier than those in the driver. The cross-correlation functions shown in the right-hand side panels of the figure confirm this anticipation behaviour, with maxima occurring at positive time shifts for all three components.

\begin{figure*}
 \includegraphics[width=0.8\textwidth]{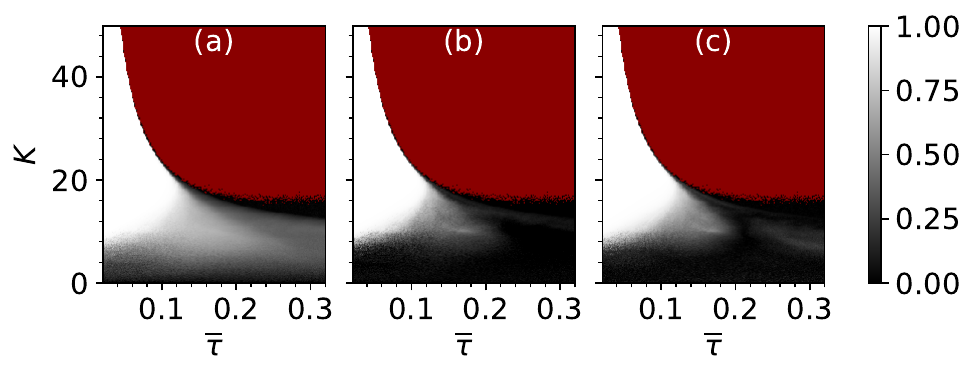}
 \caption{Maximum height of the cross-correlation function for the Lorenz system given in \Eqb{eq:Lorenz_driver}{eq:Lorenz_driven} with a uniformly distributed delay. The parameter region with approximate anticipated synchronisation is characterised by a maximum height of the cross-correlation close to one. The results obtained for the different variables are showed in the different panels: $x_1$ and $y_1$ in panel (a), $x_2$, $y_2$ in panel (b), and $x_3$ and $y_3$ in panel (c). Model parameters are $a = 10$, $b = 28$, and $c = 8/3$. The width of the delay distribution is fixed $\Delta\tau = 0.04$.}
 \label{fig:StabLorenz2Delays}
\end{figure*}
We report the results of a more systematic exploration of parameter space in Fig.~\ref{fig:StabLorenz2Delays}, indicating again, for each component, the maximum height of the correlation function for different $\overline \tau$ and $K$.  For large mean delays and/or coupling strengths the driven system diverges (red region in the figure). Outside this region of divergence we again identify approximate anticipated synchronization from values of the maximum cross-correlation function close to one. As in the R\"ossler system approximate anticipated occurs when the mean time delay and the coupling strength between the driver and the driven system are not too large.

Overall, the phase diagram of the Lorenz system closely resembles that of the R\"ossler system shown in Fig.~\ref{fig:StabRossler2Delays}. Moreover, as in the R\"ossler case, the anticipation times $\tau_i^\ast$ are found to be close to the average delay $\overline{\tau}$, with a symmetric mean absolute percentage error below $3.1\%$ for all three components. 

\section{Conclusions}\label{sec:conclusions}

We have investigated anticipated synchronization in systems with distributed delay coupling. 
We have shown that the phenomenon of AS, previously studied in coupled systems with single delay, remains robust against the introduction of distributed delay. Our analysis covers relatively simple harmonic oscillators as well as nonlinear chaotic dynamics.

For linear systems, we have obtained a complete analytical characterization of anticipated solutions under distributed delay coupling. For coupled simple damped harmonic oscillators we find AS between decaying trajectories of the driver and driven systems, provided the coupling strength or the average delay are not too large. We also find that amplification of the driver signal can occur in the driven system, i.e., the amplitude of the driven system can exceed that of the driver. 

If the driver and the driven units each consists of two coupled oscillators then each unit has multiple normal modes. Exact AS is then no longer possible, but nonetheless we detect behaviour that closely resembles AS very closely, and we again find amplification of the driver amplitude in the driven unit. We have demonstrated that the presence of multiple delay lines can enlarge the region of parameter space in which stable anticipation occurs, allowing for larger anticipation times than in for systems with single-delay coupling.

In chaotic system with distributed delay coupling we do not find exact anticipated synchronisation. However, numerical results show an approximate form of AS. Using the R\"ossler and Lorenz systems as representative examples, we find that the trajectories of the driven system closely reproduce the driver dynamics in an extended region of parameter space. The time shift between driver and driver system then typically varies along the trajectory, but nonetheless indicates anticipation with an average anticipation time close to the mean delay. The relative amplitude of the driver and driven signals also varies along trajectories, but we observed a tendency towards amplification in the driven system. 

Taken together, our results show that anticipated synchronization extends beyond single-delay coupling. Distributed delay modifies the dynamics, but it does not suppress anticipation altogether. Our findings thus broaden the theoretical framework for anticipated synchronization. This is relevant for situations in which multiple or uncertain delays are unavoidable.

As a possible extension of this work, it would be interesting to study excitable systems with distributed delay coupling for which one can use a predict-precent control strategy to prevent unwanted spikes~\cite{Ciszak_2009,Mayol_2012}. Experimental realisations of distributed delay coupling, for instance using multiple feedback loops, could provide a test of the robustness of anticipated synchronization beyond the idealized single-delay settings in previous work.

\medskip

{\bf Acknowledgments}
Partial financial support has been received from Grants PID2021-122256NB-C21/C22 and PID2024-157493NB-C21/C22 funded by MICIU/AEI/10.13039/501100011033 and by “ERDF/EU”, and the María de Maeztu Program for units of Excellence in R\&D, grant CEX2021-001164-M.


\appendix
\makeatletter
\renewcommand{\thefigure}{\Alph{section}\arabic{figure}}
\@addtoreset{figure}{section}
\@addtoreset{table}{section}
\makeatother

\section{Coupled harmonic oscillators with uniformly distributed delay}\label{app:harm_osc_uniform_delay}
In this appendix we provide the form of the analytical solution for a setup in which the driver and driven system each consists of a damped harmonic oscillator [\Eq{eq:driver_osc}], but where the delay now has a uniform distribution in the interval $\left[ \tau_\ell,\ \tau_u \right]$. 

From Eqs.~(\ref{eq:AS_solution} ,\ref{eq:beta}, \ref{eq:tstar}), the solution is found as
\begin{equation}\label{eq:SolutionY1SlaveUniformDistr}
 y_1(t) = \frac{\Omega \left( \tau_u - \tau_\ell \right)e^{b\tau^\ast}}{B}x_1(t+\tau^\ast),
\end{equation}
where
\begin{eqnarray}
 B &=&\sqrt{
 e^{2b \tau_\ell} + e^{2b \tau_u} - 2b e^{b \left( \tau_\ell + \tau_u \right)}\cos \mleft[\Omega_0 \left( \tau_u - \tau_\ell \right)\mright]
 }, \nonumber \\
 \tau^\ast &=& \Omega_0^{-1}\arctan \mleft(\frac{D}{d}\mright), \nonumber\\
 D &=&\ e^{b \tau _u} \left[b \sin\mleft(\Omega _0 \tau _u\mright)-\Omega _0 \cos\mleft(\Omega _0 \tau _u\mright)\right] \nonumber \\
 &&+e^{b \tau _\ell} \left[\Omega _0 \cos\mleft( \Omega _0\tau _\ell\mright)-b \sin\mleft( \Omega_0\tau _\ell\mright)\right], \nonumber\\
 d &=&\ e^{b \tau_u} \left[b \cos\mleft(\Omega _0 \tau_u\mright)+\Omega _0 \sin\mleft(\Omega _0 \tau _u\mright)\right]\nonumber \\
 &&-e^{b \tau _\ell}\left[b \cos\mleft(\Omega _0\tau _\ell \mright)+\Omega _0 \sin\mleft(\Omega _0\tau_\ell\mright)\right].
\end{eqnarray}

\section{Calculation of the poles of Eq.~(\ref{eq:DeltaSolutionsLaplace})}\label{app:roots}

Let us denote generically by $F(s)$ the common denominator in Eq.~(\ref{eq:DeltaSolutionsLaplace}). We need to solve $F(s)=0$ in the complex plane.

To find the roots, we use Halley’s method, which is particularly suitable here since both the first and second derivatives of $F(s)$ can be computed analytically. The method is iterative: starting from an initial guess $s_0$ for a root, the sequence is updated according to 
\begin{equation}
 s_{n+1} = s_n + \dfrac{2 F(s_n) F'(s_n)}{2 [F'(s_n)]^2 - F(s_n) F''(s_n)}.
\end{equation}
The iteration proceeds until $|s_{n+1} - s_n| < \varepsilon$, a prefixed accuracy. When this is achieved we set the converged value as $\bar{s}=s_n$.

As $F(s^*)=F(s)^*$, solutions for $s$ appear in complex conjugate pairs. Therefore, it suffices to consider only the solutions with a positive imaginary part, i.e. $s=s_R+is_I$ with $s_I\ge 0$. 

For example, in the case of equally weighted delays [Eq.~(\ref{eq:g2taus})], with $K_2 = 0$, $K_1 \equiv K$, we have
\begin{equation}\label{app:eqzeros}
 F(s) = \Omega^2 + \left(s + \frac{K}{2}(e^{-s\tau_1}+e^{-s\tau_2})\right)\left(2b + s \right).
\end{equation}
In this example, in order to find multiple roots, we use as initial guesses $s_0=s_{R,0}+is_{I,0}$, a grid of $20 \times 20$ equally spaced points in the domain $s_{R,0} \in [-20,20]$ and $s_{I,0} \in [0,20]$.  For the parameter values considered in the main text, this search region is sufficient to capture the roots with the largest real part. For each initial guess, the iteration is continued until the convergence criterion $|s_{n+1} - s_n| < 10^{-7}$ is satisfied. Since different initial guesses may converge to the same root, two converged values $\bar{s}$ and $\bar{s}'$ are regarded as identical if $|\bar{s} - \bar{s}'| < 10^{-6}$. This procedure is used to produce Fig.~\ref{fig:EigenvaluesSlaveDHO}.

In Figs.~\ref{fig:StabilityDampOsc} and \ref{fig:ComparisonTwoDelaysSingle}, we are only interested in knowing if the AS solution is stable or unstable. Therefore, all we need to determine is whether or not there are solutions of Eq.~\eqref{app:eqzeros} with a real part larger than $-b$.

To this end we use Cauchy's formula for counting the zeros of a holomorphic function $F(z)$ inside a simple closed contour~\cite{Kravanja_2000}, 
\begin{equation}\label{eq:contour}
 N = \frac{1}{2\pi i}\int_{\mathcal{C}}\dd{z}\frac{F'(z)}{F(z)}.
\end{equation}
We take a rectangular contour enclosing the area $[-b,\Gamma_1]\times[0,\Gamma_2]$, and perform the above integral numerically, using QAGS algorithm~\cite{Piessens_1983} choosing $\Gamma_1$ and $\Gamma_2$ sufficiently large. For example, in Eq.~\eqref{app:eqzeros}, we took $\Gamma_1=\Gamma_2=20$. If $N>0$ there is at least one solution of Eq.~(\ref{app:eqzeros}) with a real part larger than $-b$, and so the AS solution is unstable. Conversely, if $N=0$, the solution is stable. Similarly, we use a contour enclosing the area $[0,\Gamma_1]\times [0,\Gamma_2]$. If there is a solution inside this region the driven system will diverge.

In some cases, due to numerical errors, the numerical integration in Eq.~(\ref{eq:contour}), does not converge to an integer value. This occurs for example when there is a pole close to the contour. We then find the roots explicitly using Halley's method as explained above, with suitable initial conditions on the contour.

\bibliographystyle{elsarticle-num}
\bibliography{references_elsevier_clean}

\begin{thebibliography}{10}
\expandafter\ifx\csname url\endcsname\relax
  \def\url#1{\texttt{#1}}\fi
\expandafter\ifx\csname urlprefix\endcsname\relax\def\urlprefix{URL }\fi
\expandafter\ifx\csname href\endcsname\relax
  \def\href#1#2{#2} \def\path#1{#1}\fi

\bibitem{Voss_2000}
H.~U. Voss, Anticipating chaotic synchronization, Physical Review E 61~(5)
  (2000) 5115.
\newblock \href {https://doi.org/10.1103/physreve.61.5115}
  {\path{doi:10.1103/physreve.61.5115}}.

\bibitem{Voss_2001b}
H.~U. Voss, {Erratum: Anticipating chaotic synchronization [Physical Review E
  61, 5115 (2000)]}, Physical Review E 64~(3) (2001) 039904.
\newblock \href {https://doi.org/10.1103/physreve.64.039904}
  {\path{doi:10.1103/physreve.64.039904}}.

\bibitem{Calvo_2004}
O.~Calvo, D.~R. Chialvo, V.~M. Eguiluz, C.~Mirasso, R.~Toral, Anticipated
  synchronization: A metaphorical linear view, Chaos: An Interdisciplinary
  Journal of Nonlinear Science 14~(1) (2004) 7.
\newblock \href {https://doi.org/10.1063/1.1620991}
  {\path{doi:10.1063/1.1620991}}.

\bibitem{Masoller_2001}
C.~Masoller, D.~H. Zanette, Anticipated synchronization in coupled chaotic maps
  with delays, Physica A: Statistical Mechanics and its Applications 300~(3-4)
  (2001) 359.
\newblock \href {https://doi.org/10.1016/s0378-4371(01)00362-4}
  {\path{doi:10.1016/s0378-4371(01)00362-4}}.

\bibitem{Hernandez_Garcia_2002}
E.~Hern{\'{a}}ndez-Garc{\'{\i}}a, C.~Masoller, C.~R. Mirasso, Anticipating the
  dynamics of chaotic maps, Physics Letters A 295~(1) (2002) 39.
\newblock \href {https://doi.org/10.1016/s0375-9601(02)00147-0}
  {\path{doi:10.1016/s0375-9601(02)00147-0}}.

\bibitem{Voss_2016}
H.~U. Voss, Signal prediction by anticipatory relaxation dynamics, Physical
  Review E 93~(3) (2016) 030201(R).
\newblock \href {https://doi.org/10.1103/physreve.93.030201}
  {\path{doi:10.1103/physreve.93.030201}}.

\bibitem{Voss_2001a}
H.~U. Voss, Dynamic long-term anticipation of chaotic states, Physical Review
  Letters 87~(1) (2001) 014102.
\newblock \href {https://doi.org/10.1103/physrevlett.87.014102}
  {\path{doi:10.1103/physrevlett.87.014102}}.

\bibitem{Shahverdiev_2002}
E.~M. Shahverdiev, S.~Sivaprakasam, K.~A. Shore, Inverse anticipating chaos
  synchronization, Physical Review E 66 (2002) 017204.
\newblock \href {https://doi.org/10.1103/PhysRevE.66.017204}
  {\path{doi:10.1103/PhysRevE.66.017204}}.

\bibitem{Pyragiene_2015}
T.~Pyragien{\.e}, K.~Pyragas, Anticipating synchronization in a chain of
  chaotic oscillators with switching parameters, Physics Letters A 379~(47)
  (2015) 3084.
\newblock \href
  {https://doi.org/https://doi.org/10.1016/j.physleta.2015.10.030}
  {\path{doi:https://doi.org/10.1016/j.physleta.2015.10.030}}.

\bibitem{Ciszak_2003}
M.~Ciszak, O.~Calvo, C.~Masoller, C.~R. Mirasso, R.~Toral, Anticipating the
  response of excitable systems driven by random forcing, Phys. Rev. Lett. 90
  (2003) 204102.
\newblock \href {https://doi.org/10.1103/PhysRevLett.90.204102}
  {\path{doi:10.1103/PhysRevLett.90.204102}}.

\bibitem{voss2002real}
H.~U. Voss, Real-time anticipation of chaotic states of an electronic circuit,
  International Journal of Bifurcation and Chaos 12~(7) (2002) 1619.
\newblock \href {https://doi.org/10.1142/S0218127402005340}
  {\path{doi:10.1142/S0218127402005340}}.

\bibitem{PhysRevE.111.064206}
R.~Vidal, J.~Escalona, M.~Rivera, A.~Flores-Rosas, F.~Montoya, T.~Roy,
  P.~Parmananda, Anticipating synchronization in electrochemical systems,
  Physical Review E 111 (2025) 064206.
\newblock \href {https://doi.org/https://doi.org/10.1103/PhysRevE.111.064206}
  {\path{doi:https://doi.org/10.1103/PhysRevE.111.064206}}.

\bibitem{Masoller_2001_lasers}
C.~Masoller, Anticipation in the synchronization of chaotic semiconductor
  lasers with optical feedback, Physical Review Letters 86 (2001) 2782.
\newblock \href {https://doi.org/10.1103/PhysRevLett.86.2782}
  {\path{doi:10.1103/PhysRevLett.86.2782}}.

\bibitem{liu2002experimental}
Y.~Liu, Y.~Takiguchi, P.~Davis, T.~Aida, S.~Saito, J.~Liu, Experimental
  observation of complete chaos synchronization in semiconductor lasers,
  Applied Physics Letters 80~(23) (2002) 4306.
\newblock \href {https://doi.org/ttps://doi.org/10.1063/1.1485127}
  {\path{doi:ttps://doi.org/10.1063/1.1485127}}.

\bibitem{Sivaprakasam_2001}
S.~Sivaprakasam, E.~M. Shahverdiev, P.~S. Spencer, K.~A. Shore, Experimental
  demonstration of anticipating synchronization in chaotic semiconductor lasers
  with optical feedback, Phys. Rev. Lett. 87 (2001) 154101.
\newblock \href {https://doi.org/10.1103/PhysRevLett.87.154101}
  {\path{doi:10.1103/PhysRevLett.87.154101}}.

\bibitem{Rees_2003}
P.~Rees, P.~S. Spencer, I.~Pierce, S.~Sivaprakasam, K.~A. Shore, Anticipated
  chaos in a nonsymmetric coupled external-cavity-laser system, Physical Review
  A 68 (2003) 033818.
\newblock \href {https://doi.org/10.1103/PhysRevA.68.033818}
  {\path{doi:10.1103/PhysRevA.68.033818}}.

\bibitem{Tang_2003}
S.~Tang, J.~M. Liu, Experimental verification of anticipated and retarded
  synchronization in chaotic semiconductor lasers, Phys. Rev. Lett. 90 (2003)
  194101.
\newblock \href {https://doi.org/10.1103/PhysRevLett.90.194101}
  {\path{doi:10.1103/PhysRevLett.90.194101}}.

\bibitem{Wang_2005}
H.~J. Wang, H.~B. Huang, G.~X. Qi, Long-time anticipation of chaotic states in
  time-delay coupled ring and linear arrays, Physical Review E 71 (2005)
  015202.
\newblock \href {https://doi.org/10.1103/PhysRevE.71.015202}
  {\path{doi:10.1103/PhysRevE.71.015202}}.

\bibitem{Toral_2003}
R.~Toral, C.~R. Mirasso, C.~Masoller, M.~Ciszak, O.~Calvo, Anticipated
  synchronization in neuronal systems subject to noise, in:
  L.~Schimansky-Geier, D.~Abbott, A.~Neiman, C.~V. den Broeck (Eds.), {SPIE}
  Proceedings, Vol. 5114, {SPIE}, 2003, p. 261.
\newblock \href {https://doi.org/10.1117/12.490177}
  {\path{doi:10.1117/12.490177}}.

\bibitem{Ciszak_2004}
M.~Ciszak, R.~Toral, C.~Mirasso, Coupling and feedback effects in excitable
  systems: anticipated synchronization, Modern Physics Letters B 18~(23) (2004)
  1135.
\newblock \href {https://doi.org/10.1142/s0217984904007694}
  {\path{doi:10.1142/s0217984904007694}}.

\bibitem{Matias_2011}
F.~S. Matias, P.~V. Carelli, C.~R. Mirasso, M.~Copelli, Anticipated
  synchronization in a biologically plausible model of neuronal motifs, Phys.
  Rev. E 84 (2011) 021922.
\newblock \href {https://doi.org/10.1103/PhysRevE.84.021922}
  {\path{doi:10.1103/PhysRevE.84.021922}}.

\bibitem{Matias_2013}
F.~S. Matias, L.~L. Gollo, P.~V. Carelli, M.~Copelli, C.~R. Mirasso,
  Anticipated synchronization in neuronal motifs, {BMC} Neuroscience 14~(S1)
  (2013) P275.
\newblock \href {https://doi.org/10.1186/1471-2202-14-s1-p275}
  {\path{doi:10.1186/1471-2202-14-s1-p275}}.

\bibitem{Sausedo_Solorio_2014}
J.~Sausedo-Solorio, A.~Pisarchik, Synchronization of map-based neurons with
  memory and synaptic delay, Physics Letters A 378~(30-31) (2014) 2108.
\newblock \href {https://doi.org/10.1016/j.physleta.2014.05.003}
  {\path{doi:10.1016/j.physleta.2014.05.003}}.

\bibitem{Montani_2015}
F.~Montani, O.~A. Rosso, F.~S. Matias, S.~L. Bressler, C.~R. Mirasso, A
  symbolic information approach to determine anticipated and delayed
  synchronization in neuronal circuit models, Philosophical Transactions of the
  Royal Society A: Mathematical, Physical and Engineering Sciences 373~(2056)
  (2015) 20150110.
\newblock \href {https://doi.org/10.1098/rsta.2015.0110}
  {\path{doi:10.1098/rsta.2015.0110}}.

\bibitem{Mirasso_2017}
C.~R. Mirasso, P.~V. Carelli, T.~Pereira, F.~S. Matias, M.~Copelli, Anticipated
  and zero-lag synchronization in motifs of delay-coupled systems, Chaos: An
  Interdisciplinary Journal of Nonlinear Science 27~(11) (2017) 114305.
\newblock \href {https://doi.org/10.1063/1.5006932}
  {\path{doi:10.1063/1.5006932}}.

\bibitem{Dalla_Porta_2019}
L.~Dalla~Porta, F.~S. Matias, A.~J. dos Santos, A.~Alonso, P.~V. Carelli,
  M.~Copelli, C.~R. Mirasso, Exploring the phase-locking mechanisms yielding
  delayed and anticipated synchronization in neuronal circuits, Frontiers in
  Systems Neuroscience Volume 13 - 2019 (2019).
\newblock \href {https://doi.org/10.3389/fnsys.2019.00041}
  {\path{doi:10.3389/fnsys.2019.00041}}.

\bibitem{Stepp_2017}
N.~Stepp, M.~T. Turvey, Anticipation in manual tracking with multiple delays.,
  Journal of Experimental Psychology: Human Perception and Performance 43~(5)
  (2017) 914.
\newblock \href {https://doi.org/10.1037/xhp0000393}
  {\path{doi:10.1037/xhp0000393}}.

\bibitem{Ciszak_2009}
M.~Ciszak, C.~R. Mirasso, R.~Toral, O.~Calvo, Predict-prevent control method
  for perturbed excitable systems, Phys. Rev. E 79 (2009) 046203.
\newblock \href {https://doi.org/10.1103/PhysRevE.79.046203}
  {\path{doi:10.1103/PhysRevE.79.046203}}.

\bibitem{Mayol_2012}
C.~Mayol, C.~R. Mirasso, R.~Toral, Anticipated synchronization and the
  predict-prevent control method in the fitzhugh-nagumo model system, Phys.
  Rev. E 85 (2012) 056216.
\newblock \href {https://doi.org/10.1103/PhysRevE.85.056216}
  {\path{doi:10.1103/PhysRevE.85.056216}}.

\bibitem{Zamora-Munt_2014}
J.~Zamora-Munt, C.~R. Mirasso, R.~Toral, Suppression of deterministic and
  stochastic extreme desynchronization events using anticipated
  synchronization, Physical Review E 89 (2014) 012921.
\newblock \href {https://doi.org/10.1103/PhysRevE.89.012921}
  {\path{doi:10.1103/PhysRevE.89.012921}}.

\bibitem{Wei_2010}
H.~Wei, L.~Li, Estimating parameters by anticipating chaotic synchronization,
  Chaos: An Interdisciplinary Journal of Nonlinear Science 20~(2) (2010)
  023112.
\newblock \href {https://doi.org/10.1063/1.3429598}
  {\path{doi:10.1063/1.3429598}}.

\bibitem{Ciszak_2015}
M.~Ciszak, C.~Mayol, C.~R. Mirasso, R.~Toral, Anticipated synchronization in
  coupled complex {G}inzburg--{L}andau systems, Phys. Rev. E 92 (2015) 032911.
\newblock \href {https://doi.org/10.1103/PhysRevE.92.032911}
  {\path{doi:10.1103/PhysRevE.92.032911}}.

\bibitem{Ambika_2009}
G.~Ambika, R.~E. Amritkar, Anticipatory synchronization with variable time
  delay and reset, Phys. Rev. E 79 (2009) 056206.
\newblock \href {https://doi.org/10.1103/PhysRevE.79.056206}
  {\path{doi:10.1103/PhysRevE.79.056206}}.

\bibitem{Ciszak_2005}
M.~Ciszak, J.~M. Guti\'errez, A.~S. Cofi\~no, C.~Mirasso, R.~Toral,
  L.~Pesquera, S.~Ort\'{\i}n, Approach to predictability via anticipated
  synchronization, Phys. Rev. E 72 (2005) 046218.
\newblock \href {https://doi.org/10.1103/PhysRevE.72.046218}
  {\path{doi:10.1103/PhysRevE.72.046218}}.

\bibitem{Pyragas_2008}
K.~Pyragas, T.~Pyragien\ifmmode~\dot{e}\else \.{e}\fi{}, Coupling design for a
  long-term anticipating synchronization of chaos, Phys. Rev. E 78 (2008)
  046217.
\newblock \href {https://doi.org/10.1103/PhysRevE.78.046217}
  {\path{doi:10.1103/PhysRevE.78.046217}}.

\bibitem{Kravanja_2000}
P.~Kravanja, M.~Van~Barel, Computing the Zeros of Analytic Functions, Springer
  Berlin Heidelberg, 2000.
\newblock \href {https://doi.org/10.1007/bfb0103927}
  {\path{doi:10.1007/bfb0103927}}.

\bibitem{Piessens_1983}
R.~Piessens, E.~de~Doncker-Kapenga, C.~W. Überhuber, D.~K. Kahaner, Quadpack,
  Springer Berlin Heidelberg, 1983.
\newblock \href {https://doi.org/10.1007/978-3-642-61786-7}
  {\path{doi:10.1007/978-3-642-61786-7}}.

\end{thebibliography}

\end{document}